\documentclass[11pt,a4paper,twoside]{article}

\usepackage{epsfig}
\usepackage[latin1]{inputenc}
\def\be{\begin{equation}}
\def\ee{\end{equation}}
\def\ba{\begin{array}}
\def\bacc{\begin{array} {cc}}
\def\ea{\end{array}}
\def\bea{\begin{eqnarray}}
\def\eea{\end{eqnarray}}
\def\bd{\begin{displaymath}}
\def\ed{\end{displaymath}}
\def\a{\alpha}
\def\b{\beta}

\begin{document}

\begin{center}

{\Large\bf Stability and Negative Tensions \\ in 6D Brane Worlds}

\vspace{1cm}

{\large S. L. Parameswaran$^a$\footnote{Email: param@sissa.it}, S.
Randjbar-Daemi$^b$\footnote{Email: seif@ictp.trieste.it} and
A. Salvio$^c$\footnote{Email: alberto.salvio@epfl.ch}}\\

\vspace{.6cm}

{\it {$^a$ Scuola Internazionale Superiore di Studi Avanzati,\\
Via Beirut 2-4, 34014 Trieste, Italy}}

\vspace{.4cm}

{\it {$^b$ International Center for Theoretical Physics, \\Strada
Costiera 11, 34014 Trieste, Italy}}

\vspace{.4cm}

{\it {$^c$ Institut de Th\'eorie des Ph\'enom\`enes Physiques,
\\EPFL, CH-1015 Lausanne, Switzerland}}

\end{center}

\vspace{1cm}

\begin{abstract}

We investigate the dynamical stability of warped, axially symmetric compactifications in anomaly free 6D gauged supergravity.  The solutions have conical defects, which we source by 3-branes placed on orbifold fixed points, and a smooth limit to the classic sphere-monopole compactification.  Like for the sphere, the extra fields that are generically required by anomaly freedom are especially relevant for stability.  With positive tension branes only, there is a strict stability criterion (identical to the sphere case) on the charges present under the monopole background.  Thus brane world models with positive tensions can be embedded into anomaly free theories in only a few ways.  Meanwhile, surprisingly, in the presence of a negative tension brane the stability criteria can be relaxed.  We also describe in detail the geometries induced by negative tension codimension two branes.

\end{abstract}

\newpage

\tableofcontents

\newpage

\section{Introduction}
Much has been written about brane world solutions to chiral gauged 6D supergravity.  Gibbons, G\"uven and Pope (GGP) \cite{GGP} found a wide class of such solutions, with 4D Poincar\'e symmetry and axial symmetry in the transverse dimensions.  Surrounding work focusing on the brane world interpretation of these backgrounds was made in \cite{warpedbranes, generalsoln}.  Other classes of solutions have also been found, including those which break the axial symmetry \cite{nonaxial}, activate the hyperscalar fields \cite{superswirl}, have 4D de Sitter/anti de Sitter slicings \cite{desitter} or time-dependent behaviour \cite{scaling, time2, time3}.  The model additionally admits string solutions with dyonic charges \cite{dyonic}.  All these backgrounds are interesting because, among other things, they provide a framework in which to build cosmological models (for a review see \cite{sled}).

In order for these backgrounds to be phenomenologically relevant, however, we would like them to be stable against small perturbations\footnote{Or at least not to exhibit runaways that are too fast.}.  So far, the GGP solutions have proven to be classically marginally stable, despite the fact that they all but one ({\it i.e.} the sphere-monopole limit) break supersymmetry (see \cite{kicking}, and also \cite{leepapa, massgaps, gravitino}).  Meanwhile, we might also want to consider the 6D theory to be a low energy approximation to some consistent theory of quantum gravity, such as string theory.  If this is the case we must insist on certain consistency constraints, and in particular, since the theory is chiral, we must insist on anomaly freedom.

In general, chiral 6D supergravity suffers from a breakdown of local symmetries due to gravitational, gauge and mixed anomalies.  For certain gauge groups and hypermultiplet representations these anomalies can be cancelled via a Green-Schwarz mechanism \cite{seifanomaly}.  This is entirely analogous to what happens in 10D, where the anomalies cancel only for a few models, namely those with gauge groups: $SO(32)$, $E_8 \times E_8$, $E_8 \times U(1)^{248}$ and $U(1)^{496}$.  In 6D the consistency constraints are weaker, and by now a number of anomaly free models have been discovered  \cite{seifanomaly,Avramis:2005qt,Avramis:2005hc,Suzuki:2005vu}.  In Table \ref{anomalyfree} we present three of the known examples which have a large enough gauge group to include the Standard Model of Particle Physics.

The structure of these anomaly free models seems suggestive that some of them may indeed be somehow related to critical string theory or M-theory \cite{stringyorigin}.  Meanwhile, with regards to the stability of the brane world compactifications, the extra degrees of freedom required for anomaly cancellation cannot be ignored.   Marginal stability was affirmed in \cite{kicking}, for the GGP solution in the Salam-Sezgin model, which has just an Abelian $U(1)_R$ gauge group and no hypermatter.  We now ask if there is a similar dynamics in anomaly free models, which have larger field contents.

To this end, we may draw some lessons from the old literature on sphere compactifications, which are supported by monopole backgrounds \cite{Salam:1984cj}.  The stability of sphere compactifications for nonsupersymmetric theories was studied in \cite{nonsusysphere, seifinstabilities}, and for anomaly free supergravity theories in \cite{seifanomaly,Avramis:2005qt,Avramis:2005hc}.  Whilst the models are stable in the presence of just a Maxwell gauge group, for Yang-Mills theories a tachyonic instability is generically found in the scalars descending from the gauge fields and charged under the $U(1)$ monopole background \cite{seifinstabilities}.  For example, it turns out that only one of the anomaly free models presented in Table \ref{anomalyfree} has a stable sphere compactification if the $U(1)$ monopole is embedded in a non-Abelian factor of the gauge group: the $E_7 \times E_6 \times U(1)_R$ model with the monopole embedded in $E_6$ \cite{seifanomaly,Avramis:2005hc,alberto}.

\begin{table}[top]
\begin{center}
\begin{tabular}{|l|l|}
\hline  Gauge Group & Hyperino Representation  \\ \hline
 $E_7 \times E_6 \times U(1)_R$ & $({\bf 912},{\bf
1})_0$ \\
 $E_7 \times G_2 \times U(1)_R$ & $({\bf 56},{\bf 14})_0$  \\
$F_4 \times Sp(9) \times U(1)_R$ & $({\bf 52},{\bf 18})_0$
 \\ \hline
\end{tabular}
\end{center}\caption{\footnotesize Some examples of anomaly free models with gauge groups containing $SU(3)\times SU(2) \times U(1)$ \cite{seifanomaly,Avramis:2005qt,Avramis:2005hc}.  There are also many other models, including {\it e.g.} one with just an Abelian $U(1)_R$ gauge group, and the anomalies cancelled by 245 neutral hypermultiplets \cite{Suzuki:2005vu}. \label{anomalyfree}}
\end{table}

Our main focus in the present work is then on the scalar perturbations of the gauge fields charged under the $U(1)$ monopole background, as possible sources of instability.  We analytically solve the linearized dynamics of these fields, and in particular derive their full Kaluza-Klein mass spectra.  In this way we are able to identify some conditions for stability, and we observe the previous behaviour as well as some surprises.

Our results can be summarised as follows.  Conical-GGP solutions which incorporate only positive tension brane sources are stable only for very special matter contents and monopole embeddings.  Specifically, the stability criteria observed in the sphere limit persists for these more general solutions.  The sphere's stability criteria is also sufficient to ensure stability for conical-GGP solutions in the presence of negative tensions (placed on orbifold fixed points).  However, remarkably, we find that negative tension branes can relax these conditions, and render unstable sphere compactifications stable.

Let us end with an overview of the paper.  In Section \ref{S:theory} we briefly review the theory and its warped, axially symmetric brane world solutions.  We also discuss some physical aspects of the background, in particular emphasising that the geometry induced by the backreaction of negative tension branes is well-defined.  Then, in Section \ref{S:scalars} we classify all the scalar perturbations present in the model and identify how the various sectors decouple.  We argue that possible tachyonic instabilities should lie in the scalar fluctuations of the gauge fields orthogonal to the background monopole in the Lie algebra of the gauge group.    Therefore, we turn in Section \ref{S:oursector} to the Kaluza-Klein mass spectra of these fields, and dedicate Section \ref{S:stability} to the consequences of these spectra for the stability.  We end with some conclusions, and leave for the appendices some details on the algebra.

\section{6D Supergravity and its Axially Symmetric Solutions} \label{S:theory}

In order to fix our conventions, we begin by reviewing the 6D chiral gauged supergravity and brane world solutions that interest us.  Then, in the following subsection, we will collect some details about the background geometry and topology which will later prove to be important.

\subsection{The theory and solution}
We consider 6D supergravity with a
general matter content, whose gauge group $\mathcal{G}$ is a product of simple groups that include a $U(1)_R$ gauged R-symmetry.  For example
we could take the anomaly free group
$\mathcal{G}=E_7 \times E_6 \times U(1)_R$, under which the
hyperinos are charged as $\Psi \sim ({\bf 912},{\bf 1})_0$ \cite{seifanomaly}.
The bosonic action takes the form\footnote{We choose signature $(-,+,..,+)$, and define $R=G^{MN}(\partial_P \Gamma^P_{MN}-\partial_M \Gamma^P_{PM} + \dots)$.  The index $M$ runs over $0,1,..,5$.  For fermionic terms see
\cite{Nishino:1984gk}.} \cite{Nishino:1984gk}
\bea S_B &=&\int d^6 X
\sqrt{-G}\left[\frac{1}{\kappa^2}R-\frac{1}{4}\partial_M\sigma\partial^M\sigma
-\frac{1}{4}e^{\kappa\sigma/2}Tr\left(F_{MN} F^{MN} \right)
\right.\nonumber \\
&&\qquad \left.-\frac{1}{12}e^{\kappa\sigma}H_{MNP} H^{MNP} -g_{\a
\b}(\Phi)D_M\Phi^{\a}D^M\Phi^{\beta} -\frac{8}{\kappa^4}e^{-\kappa
\sigma/2}v(\Phi)\right],
\label{SB} \eea
where $\kappa$ represents the 6D Planck scale and $g$ is the gauge
coupling constant, which in fact represents a collection of
independent gauge couplings including that of the $U(1)_R$
subgroup, $g_1$.  The field $\sigma$ is the dilaton, $F_{MN}$ is
the field strength of the gauge field, ${\cal A}_M$, and $H_{MNP}$
is the Kalb-Ramond field strength, which contains a Chern-Simons
coupling as follows\footnote{We define the cross-product as
$({\cal A}_M \times {\cal A}_N)^{\hat I} = f^{\hat I\hat J\hat K}
{\cal A}_M^{\hat J}{\cal A}_N^{\hat K}$, with $f^{\hat I\hat J\hat
K}$ the structure constants of ${\mathcal G}$.  The index $\hat I$
runs over the full Lie algebra of ${\mathcal G}$, and later we
will use $I$ to label those directions orthogonal to that of the
$U(1)$ monopole.}: \be H_{MNP} = \partial_M B_{NP} +  Tr
\left[F_{MN}{\cal A}_P - \frac{g}{3}{\cal A}_M\left({\cal A}_N
\times {\cal A}_P \right)\right] + perms. \label{HKR}\ee The
metric $g_{\a\b}(\Phi)$ is on the target manifold of the
hyperscalars, and here the index $\a$ runs over all the
hyperscalars.  The dependence of the scalar potential on
$\Phi^{\a}$ is such that its minimum is at $\Phi^{\a} = 0$, where
it takes a positive-definite value, $v(0) = g_1^2$, due to the
R-symmetry gauging \cite{NS2, dyonic}.

We refer to \cite{massgaps} for the equations of motion that follow from (\ref{SB}).  A general class of configurations with 4D
Poincar\'e symmetry and axial symmetry in the transverse dimensions, is:
\bea && ds^2=G_{MN}dX^M dX^N=e^{A(\rho)}\eta_{\mu
\nu}dx^{\mu}dx^{\nu}+d\rho^2+e^{B(\rho)}d\varphi^2,\nonumber\\
&& \qquad \mathcal{A}=\mathcal{A}_{\varphi}(\rho) Q d\varphi, \qquad \sigma =\sigma(\rho),\nonumber \\
&& \qquad \qquad H_{MNP} = 0 \, , \qquad
\Phi^{\alpha} = 0 \, ,
\label{axisymmetric} \eea
with $0\leq \rho \leq \overline{\rho}$ and $0\leq \varphi<2\pi.$
Here $\mu,\nu=0,1,2,3$ and $Q$ is a generator
of a $U(1)$ subgroup of a simple factor of $\mathcal{G}$, satisfying
$Tr\left(Q^2\right)=1$.

In the following we shall also use the radial coordinate defined by
\be \xi(\rho)\equiv \int_0^{\rho}d\rho' e^{-A(\rho')/2},
\label{theta}\ee
whose range is $0\leq \xi \leq \overline{\xi}$. In this frame the metric reads
\be ds^2=e^{A(\xi)}\left(\eta_{\mu \nu}dx^{\mu}dx^{\nu}+d\xi^2\right)
+e^{B(\xi)}d\varphi^2 \, .\ee

Given the above ansatz, the general solution has been found by GGP \cite{GGP}.
We will focus on a subset of this general solution, namely that
which contains singularities no worse than conical. The explicit conical-GGP solution\footnote{The coordinate
$\xi$ is related to the coordinate $r$ in \cite{GGP} by
$r=r_0\cot(\xi/r_0)$.} is then \cite{GGP}:
\bea e^A&=&e^{\kappa \sigma/2}=\sqrt{\frac{f_1}{f_0}}, \quad e^B=\alpha^2
e^A\frac{r_0^2\cot^2(\xi/r_0)}{f_1^2},\nonumber\\
\mathcal{A}&=&-\frac{4\alpha}{q\kappa f_1}\, Q\,
d\varphi,\label{GGPsolution}\eea
where $q$ and $\alpha$ are generic real numbers.  Also,
\be f_0\equiv 1+\cot^2\left(\frac{\xi}{r_0}\right), \quad f_1 \equiv
1+\frac{r_0^2}{r_1^2}\cot^2\left(\frac{\xi}{r_0}\right), \label{GGPsolution2}\ee
with $r_0^2\equiv \kappa ^2/(2g_1^2)$, $r_1^2\equiv 8/q^2$.

The conical-GGP configuration is, however, a
solution to the equations of motion only outside the points $\xi = 0$ or $\xi = \overline{\xi}\equiv \pi r_0/2$.  This is because as $\xi \rightarrow 0$ or $\xi \rightarrow \overline{\xi}$, the metric tends to that of a cone, with respective deficit angles
\be \delta = 2\pi \left(1-|\alpha| \, \frac{r_1^2}{r_0^2}\right)
\quad \mbox{and}\quad
\overline{\delta}=2\pi\left(1-|\alpha| \right) \, , \label{deltadeltabar}\ee
and corresponding delta-function behaviours in the Ricci scalar. Note that $\alpha$ appears in the deficit angles only through its modulus, since the metric (\ref{GGPsolution}) depends only on the square of $\alpha$, and so is insensitive to its sign.  In order to promote the solution to a global one, we introduce two 3-brane sources into the system, each with action:
\be \label{braneaction}
S_b = -T \int d^4y \sqrt{-det\left(G_{MN} \partial_{\alpha} Y^M \partial_{\beta} Y^N\right)}\, ,
\ee
where $Y^M(y^{\alpha})$ are the brane embedding fields, $y^{\alpha}$ are the worldvolume coordinates, $\alpha = 0, \dots, 3$ and the tensions are respectively \cite{chenlutyponton}
\be T=2\delta/\kappa^2
\quad \mbox{and} \quad T=2\overline{\delta}/\kappa^2.\label{Tdelta}\ee
In this way we arrive at a warped codimension-two brane world construction, in which the 3-branes can localize bulk fields \cite{massgaps} and/or support 4D fields in such a way as to realize the Standard Model of Particle Physics.

Finally, we note that one can obtain the ``rugby ball'' compactification \cite{Carroll:2003db} simply by setting $r_0=r_1$.
In this case the background value of the dilaton is zero, and therefore the
stability analysis that we are going to present will also be applicable to the rugby ball solution of non-supersymmetric 6D Einstein-Yang-Mills models.  Moreover, we can smoothly retrieve the sphere compactification by taking $r_1 \rightarrow r_0$ and $\alpha \rightarrow 1$.

\subsection{Geometry and topology} \label{S:geometry}

\begin{figure}
\centering
\epsfig{file=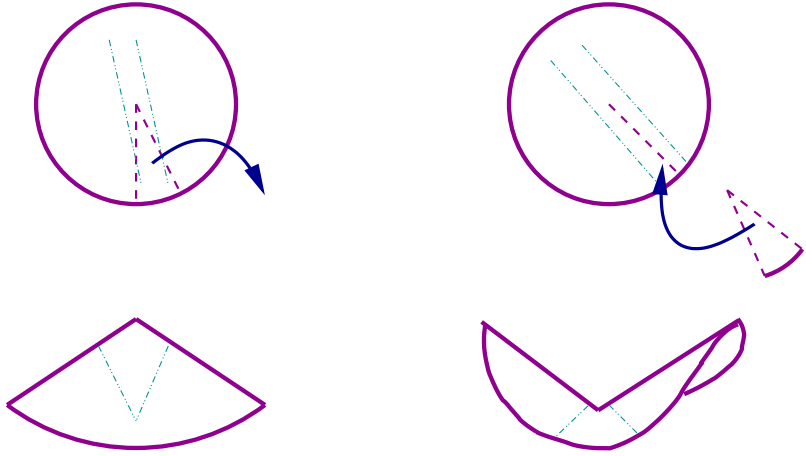,width=0.7\linewidth,clip=}
\caption{\small Construction of a cone and saddle-cone, respectively, by splicing out and in, respectively, a wedge starting from the flat disk.  For the cone, lines that were parallel on the disk remain parallel until they pass on either side of the apex, when they begin to converge.  For the saddle-cone they diverge once passing the apex.  Notice that although the saddle-cone appears to break the axial-symmetry, this is only an effect of the embedding into 3D.  A 2D being would indeed observe the axial symmetry.}\label{fig:geodesics}
\end{figure}

We continue our discussion on the background configuration by considering in more detail its geometry.  In particular, it is interesting to note that the parameters appearing in the deficit angles, $\alpha$ and $r_1$,
are not fixed by the EOM but rather represent moduli.  However, from (\ref{deltadeltabar}), we can see that the deficit angles are both bounded from above by $2\pi$. This becomes an upper bound on the brane tensions that can be described by the conical-GGP solution,
because of the matching conditions in (\ref{Tdelta}).

Meanwhile it is also clear that the deficit angles can take
arbitrary large negative values. We emphasise here that manifolds
with negative deficit angles are perfectly well defined, and can
even be made at home with a piece of paper and a pair of scissors.
Take for example the simplest case of a cone.  A cone with
positive deficit angle is obtained by splicing a wedge out of a
flat disk and glueing together the edges.  Similarly, a cone with
negative deficit angle is obtained by splicing a wedge {\it into}
a flat disk\footnote{As this manuscript was being prepared Ref.
\cite{Gogberashvili:2007gg} appeared, baptizing manifolds with
spherical topology and negative deficit angles as
``Apple-like''.}. The result will be a manifold which is flat
everywhere apart from at the apex of the cone. Lines that were
parallel on the disk remain parallel until they pass on either
side of the apex, after which they will begin to diverge, just as
for the standard cone they would converge (see Figure
\ref{fig:geodesics}). Moreover, there is no lower bound on the
deficit angle.  For example, one could imagine adding $\pi/3$
wedges successively to the flat disk {\it ad infinitum}.

\begin{figure}
\centering
\epsfig{file=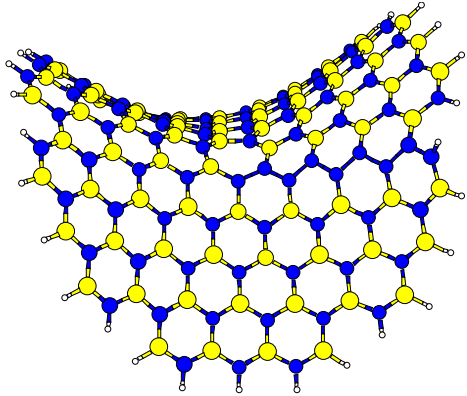,width=0.4\linewidth,clip=}
\caption{\small A nanocone with a disclination angle $-\pi/3$.  Reproduced from Ref. \cite{cond-mat} by kind permission of the authors.}\label{fig:nanocone}
\end{figure}

Discretized versions of manifolds with negative deficit angles can be found in the solid-state literature on Carbon Nanostructures, in which nanocones with negative disclination angles are appropriately named ``saddle-cones'' \cite{cond-mat}.  These provide another nice way to visualize the geometries, including those with, say, deficit angle less than $-2\pi$.  Referring to
Figure \ref{fig:nanocone}, we observe that beginning with a flat planar lattice of regular hexagons, and splicing in a wedge of $\pi/3$, one ends with another lattice of hexagons, now taking the form of a (flat) saddle, and with the central hexagon replaced by a heptagon.  A deficit angle of $-\pi$ would correspond to a central enneagon, one of $-2\pi$ to a dodecagon, and so on.

On the other hand, we should note from Eq. (\ref{Tdelta}) that in our scenario the negative deficit angles are sourced by negative tension branes.  It is well known that negative tension branes generically suffer from classical and quantum instabilities \cite{negativetensions}.  We will return to this issue in the following section.

Having understood the geometry of the model, we must also take care of its topology.  The Euler number of the internal manifold is two, indicating the topology of a sphere.  This can be seen from direct calculation, as well as by showing that the manifold can be covered by two holomorphic coordinate patches, which are related at the intersection by $z = 1/\zeta$ (see below).  We must therefore take care to ensure that the gauge field background is well-defined as $\xi \rightarrow 0$ and $\xi \rightarrow \overline{\xi}$.  Indeed the expression for the gauge field background in eq. (\ref{GGPsolution}) is
well-defined in the limit $\xi\rightarrow 0$, but not as $\xi \rightarrow
\overline{\xi}$.  We must therefore use a different patch to describe the
$\xi={\overline{\xi}}$ limit, and in the overlap this must be related to the patch including
$\xi=0$ by a single-valued gauge transformation.  This leads to a Dirac
quantization condition, which for a field interacting with
$\mathcal{A}$ through a charge $e$ gives
\be -e \, \frac{4\alpha {\overline g}}{\kappa q}= -e \,\alpha
\frac{r_1}{r_0} \frac{{\overline g}}{g_1} = N \, ,
\label{DiracQ}\ee
where $N$ is an integer that is called monopole number and
${\overline g}$ is the gauge coupling constant corresponding to
the background gauge field. For example, if ${\mathcal A}$ lies in
$U(1)_R$, then ${\overline g}=g_1$.  Generally different fields
have different charges $e^i$, which correspond to several monopole
numbers  $N^i$; as we shall see this aspect is important in the
stability analysis of the conical-GGP solutions. The charges $e^i$
can be computed once we have selected the gauge group, since they
are eigenvalues of the generator $Q$.

As has long and often been noted, the Dirac quantization leads to a constraint, relating the tensions of the two 3-branes which can be described by the conical-GGP solution and the bulk gauge couplings \cite{GGP}.  For example, embedding the monopole in the $U(1)_R$ gauge sector requires the presence of at least one negative tension brane \cite{GGP, generalsoln}.

\section{The Scalar Fluctuations} \label{S:scalars}
We now consider scalar fluctuations about the brane world solution, with the aim of studying its stability.  The bulk perturbations which are scalars from the brane point of view can be written as:
\be
\left\{ \delta G_{\mu}^{\mu}, \delta G_{\rho\rho}, \delta G_{\varphi\varphi}, \delta G_{\rho\varphi}, \delta \sigma, \delta \zeta, \delta B_{\rho\varphi}, \delta {\cal A}_{\rho}, \delta {\cal A}_{\varphi}, \delta \Phi \right\} \, .
\label{perts}
\ee
Here, $\delta \zeta$ is the four-dimensional dual of the fluctuation in the Kalb-Ramond field $\delta B_{\mu\nu}$.
Moreover, the presence of branes introduces additional dynamical degrees of freedom, which are the brane-bending modes corresponding to fluctuations in their position in six-dimensional spacetime:
\be
\left\{\delta Y^{M} \right\} \, .
\ee
Let us now discuss these various fluctuations in turn.

\subsection{Brane bending sector}
We begin by making some general comments.  By choosing the so-called static gauge for the worldvolume coordinates, $\partial_\alpha Y^{\mu}=\delta^{\mu}_{\alpha}$, we see that each brane carries two physical fields which correspond to their bending in the transverse dimensions:
\be
\left\{\delta Y^{\rho}(x), \delta Y^{\varphi}(x)\right\} \, .
\ee
Notice that the fields are four dimensional, and so do not lead to a Kaluza-Klein tower.  The case of the unwarped rugby ball is, at least in part, simple.  Two of the fields parametrise the motion that keeps the branes at antipodal points of the sphere, and preserves the axial symmetry of the background.  They must be massless modes.  Indeed, since the branes break the $SU(2)$ isometry group of the sphere down to the $U(1)$ axial symmetry, these fields provide the massless Goldstone bosons to be eaten by two of the $SU(2)$ massless graviphotons.  Less can be said about the remaining two brane bending fields, which describe the relative motion of the branes and which break the axial symmetry\footnote{We thank Cliff Burgess for an email communication on these issues.}.  The warped model may also be more complicated, since there the $SU(2)$ isometry of the sphere is broken not only by the brane defects but also by the geometry away from the branes.

In general we can expect the brane bending modes to mix with the bulk metric fluctuations at the bilinear level.  Moreover, it is well known that negative tension branes generically have ghost kinetic terms for their bending modes, and thus lead to energies unbounded from below and instability.  For these reasons we choose to place the branes at orbifold fixed points in such a way that they are not free to fluctuate.

To find an orbifold projection that serves this purpose, we need a global description of the internal manifold, and in particular one in which both the brane positions can be well described.   To this end, we cover the manifold with two coordinate patches.  First we define the complex coordinates $z, {\bar z}$, with:
\be
 z = \eta \, e^{i\varphi} \qquad
 \mbox{where}\,\,\,\,\,\,\eta=e^{\int_{\rho_0}^{\rho} e^{-B/2}}
\ee
for some arbitrary $\rho_0$.  The metric on the internal manifold becomes:
\be
ds^2 = d\rho^2 + e^B d\varphi^2 = \frac{e^{B\left(\rho(z\bar z)\right)}}{z\bar z} \left(dz \, d\bar z \right) \,.
\ee
Using the behaviour of $e^{B}$ as $\rho \rightarrow 0$ ($e^B \rightarrow (\alpha \, r_1^2/r_0^2)^2 \rho^2$), one can show that the brane at $\rho=0$ is now well described by the single point $z=0$.  Meanwhile, in analogy with standard stereographic coordinates, the point $\rho=\bar\rho$ cannot be covered by $z,\bar z$.  We therefore use a different coordinate patch to describe the brane there, defined by $\zeta, \bar\zeta$ with:
\be
 \zeta = \frac{1}{\eta} \, e^{-i\varphi} \, .
\ee
In this patch the brane at $\rho=\bar\rho$ is well-defined at $\zeta=0$, and it is the brane at $\rho=0$ which is not covered.  In the overlap, the two coordinate systems are related by $\zeta = 1/z$, $\bar\zeta = 1/\bar z$.  Notice that this confirms that the internal space is conformally $CP^1$.

Now one can immediately see that the brane positions are the fixed points under the orbifold identification
\be
z \leftrightarrow -z \, ,\, \, \, \, \, \zeta \leftrightarrow -\zeta
\ee
The brane fluctuations, $\delta Y^z(x), \delta Y^{\bar z}(x)$ and $\delta Y^{\zeta}(x), \delta Y^{\bar \zeta}(x)$, being odd under the orbifold action, are thus projected out.

\subsection{Hyperscalars}
It is straightforward to see that the hyperscalars do not mix with
the other sectors at the level of the bilinear action.  Moreover, we can quite directly conclude they cannot give rise to tachyonic instabilities: this is a consequence of the fact that the potential $v(\Phi)$ has a global minimum at $\Phi=0$, and also that the metric $g_{\a\b}(\Phi)$ is positive definite.  Therefore, the contributions from the 6D potential and the 2D Laplacian to the mass-squareds in the Kaluza-Klein mass spectrum are both positive.

\subsection{Scalars from the gauge fields}
The scalar fluctuations descending from the gauge fields
can be divided into two separate classes.  First there are the
fluctuations:
\be
\left\{\delta {\cal A}_{\rho} \, Q, \delta {\cal
A}_{\varphi}\,Q\right\} \,,  \label{||}
\ee
with the gauge group generator $Q$
corresponding to that of the background monopole.  These will be included in the discussion of the following subsection.  Second, there are
the fluctuations
\be\left\{\delta {\cal A}_{\rho}^I \, T^I, \delta {\cal A}_{\varphi}^I \, T^I\right\},\label{oursector}\ee
with $T^I$ being the generators orthogonal to $Q$, that is $Tr(T^IQ)=0$.
At the bilinear level the second group does not mix with the first, and nor with any of the other sectors in (\ref{perts}).  This is a consequence of the form of the action (\ref{SB}) and that of the Kalb-Ramond field strength (\ref{HKR}).  We will therefore return to the fields (\ref{oursector}), which are the main focus in the present paper, in Section \ref{S:oursector}.  The possible instabilities are generally lurking in this sector.

\subsection{Salam-Sezgin sector} \label{symmetries}

The remaining fields are those which also correspond to the minimal Salam-Sezgin model, that is with just one $U(1)_R$ gauge multiplet:
\be
\left\{ \delta G_{\mu}^{\mu}, \delta G_{\rho\rho}, \delta G_{\varphi\varphi},
\delta G_{\rho\varphi}, \delta \sigma, \delta \zeta, \delta
B_{\rho\varphi}, \delta {\cal A}_{\rho}\,Q, \delta {\cal
A}_{\varphi}\,Q \right\}\,. \label{cliffsector}
\ee
Let us here recall the long-known result that the Salam-Sezgin sphere model is marginally
stable, with two and only two massless scalar modes, and a Kaluza-Klein tower of heavy
positive mass-squared modes \cite{seifnote, seifanomaly,toykklt,alberto}.  One of the massless modes corresponds to the spontaneous breaking of the global classical scaling symmetry\footnote{The EOMs are invariant under the constant rescaling $G_{MN} \rightarrow \lambda \, G_{MN}$ and $e^{\kappa\sigma/2} \rightarrow \lambda \, e^{\kappa\sigma/2}$.  Note that this is only a classical symmetry because the action rescales as $S_B \rightarrow \lambda^2 \, S_B$.}.  The other is guaranteed by the unbroken Kalb-Ramond gauge symmetry\footnote{This symmetry acts as $B\rightarrow B+d\Lambda$, where $B$ is the two-form with components $B_{MN}$, and $\Lambda$ a one-form gauge parameter.  Then the components $B_{\mu\nu}$ are dual to a massless 4D scalar.  Meanwhile, the directions $B_{\rho\varphi}$ provide the Goldstone boson that is to be eaten by the massive $U(1)$ gauge field in the direction of the monopole.}.  Since the massless modes are protected by symmetry arguments, we can argue that small deformations of the sphere solution to the conical-GGP solution must remain marginally stable.

Moreover, Ref.~\cite{kicking} analysed explicitly a subsector of (\ref{cliffsector}), namely the axially-symmetric perturbations corresponding to those members that are even under a certain parity symmetry.  By imposing that the perturbations preserve the conical singularities,\footnote{Actually, the general GGP solutions with worse than conical singularities were also considered in \cite{kicking}.} a single massless mode was found, corresponding to the classical scaling symmetry enjoyed by the field equations\footnote{The second massless mode can be expected amongst the perturbations that are odd under the parity symmetry.}.  All other modes were shown to have positive squared-masses.

\section{Linear Analysis for Scalar Fluctuations of the Gauge Fields} \label{S:oursector}
We will now complete the stability analysis for brane world compactifications in anomaly free models, by considering
the final sector $\left\{\delta {\cal A}_{\rho}^I \, T^I, \delta {\cal A}_{\varphi}^I \, T^I\right\}$.  In
fact, these fields are of particular interest. Indeed, it has long
been known that this sector -- and only this sector -- can in general contain tachyonic modes
in its Kaluza-Klein spectra for the sphere
compactification with a monopole
background \cite{seifinstabilities}. It is
therefore interesting to ask what happens to these tachyons if one
considers the conical-GGP configuration, which as we have seen is a warped deformation of the sphere compactification.

As an example, we could consider the anomaly free model of $E_7
\times E_6 \times U(1)_R$, with the monopole embedded in $E_6$.  In
this case, the low-energy gauge group is\footnote{In the sphere limit the $U(1)_{KK}$ is promoted to $SU(2)_{KK}$.}: $E_7 \times SO(10) \times
U(1)_R \times U(1)_{KK}$, and the fluctuations covered by our analysis are two sets
of scalars transforming under $E_7 \times SO(10) \times
U(1)_R$ as $({\bf 133},{\bf 1})_0+({\bf 1},{\bf 45})_0 + ({\bf
1},{\bf 16})_0 + ({\bf
1},{\bf {\overline{16}}})_0 + ({\bf 1},{\bf 1})_0$, and in various representations of $U(1)_{KK}$ depending on the monopole number.  In order to keep the analysis of the present sector as general as possible, we choose here not to impose the orbifold boundary conditions discussed above.  These would of course only project out some of the modes in the full spectra derived below, and so would not introduce new instabilities.

\subsection{Bilinear action}
Let us then consider the bilinear action for the fluctuations
\bea
V_{\rho} &=& V_{\rho}^I T^I \equiv \delta {\cal A}^I_{\rho}T^I \,,\nonumber \\
V_{\varphi} &=& V_{\varphi}^I T^I \equiv \delta {\cal A}^I_{\varphi}
T^I \eea around the background (\ref{axisymmetric}). Since $T^I$ is
orthogonal to $Q$, the perturbed action simplifies considerably,
having contributions only from the gauge kinetic term in (\ref{SB}).
After fixing to the light-cone gauge (see reference \cite{lightcone}
and Appendix \ref{A:gaugefixing} for details), the result can be
written as\footnote{In (\ref{b2action}) the fluctuation
fields $V_i$ have been normalized in a way that the canonical factor
$-1/2$ appears in front of the kinetic terms.}: \bea
S_{2}(V,V) &=& -\frac12 \int d^6X \sqrt{-\hat G} \, Tr \, \left[\partial_{\mu} V_i \partial^{\mu} V^i + D_i V_j D^i V^j - 2 (\partial_r \hat A)^2 V_r^2  \right. \nonumber \\
&& \phantom{000000000}  \left. - 2 (\partial_r \hat A) V_r D_iV^i
 + \hat R_{ij}V^iV^j + 2 \overline{g}\,F_{ij}V^i \times V^j \right] \, \label{b2action}
\eea
where we have introduced $dr \equiv e^{\kappa\sigma/4} d\rho$,
and the indices $i,j$ run over $r, \varphi$.  Also for compactness, we have defined:
\bea
&& \hat A \equiv A + \phi \, , \qquad \hat B \equiv B + \phi \, , \qquad
\phi \equiv \kappa\sigma/2 \, \nonumber\\
&& \hat G_{MN} dX^M dX^M \equiv e^{\hat A} \eta_{\mu\nu} dx^{\mu}
dx^{\nu} + dr^2 + e^{\hat B} d\varphi^2 \, , \eea and all indices
are raised and lowered with the background metric $\hat G_{MN}$.
$\hat R_{ij}$ are the internal components of the Ricci-tensor
defined from the metric $\hat G_{MN}$, and $F_{ij}$ refers to
the background field strength.

Moreover, recall that the covariant derivative in general includes
the gauge field background.  In particular, we have \be
D_{\varphi} V_j = {\nabla}_{\varphi} V_j - i\overline{g}\, {\cal
A}_{\varphi}[Q, V_j] \ee with ${\nabla}_{\varphi}$ the Lorentz
covariant derivative.  Below we will choose a basis of generators
such that: \be [Q,T^I] = e^I T^I \, , \ee which means that in
general they will not be Hermitian. However, we choose the
normalization $Tr (T^{I\dagger} T^J) = \delta^{IJ}$, and also
define $[T^I,T^J] = i f^{IJK} T^K$.  Also, $e^I$ is the
corresponding charge under the $U(1)$ monopole. For example, for
the $E_7 \times E_6 \times U(1)$ model, we have $e^I \neq 0$ for
the ${\bf 16}$ and ${\bf {\overline{16}}}$. The Dirac quantization
condition (\ref{DiracQ}) then gives $-e^I 4 \alpha {\overline g}
/(\kappa q) = N^I$.  In the following, we suppress the index $I$.

Finally, since our internal space is topologically $S^2$, we shall
impose that the fluctuations are periodic functions of $\varphi$.
Therefore, we can apply the following Fourier decomposition: \be
V_j(X) = \sum_m V_{jm}(x,r) e^{im\varphi} \label{fourier} \ee with
$m$ an integer, $-\infty < m < \infty$.

\subsection{The equations of motion and boundary conditions}
Next we vary the above action with respect to $V_r$ and $V_{\varphi}$,
perform the Fourier decompositions (\ref{fourier}), and
project onto the Fourier number $m$.  After a long but standard calculation we eventually obtain the following
coupled equations of motion:
\bea
e^{-\hat A} M_{m}^2 V_{rm} &=& -\partial_r^2 V_{rm}-\left( 2 \partial_r \hat A + \frac12\partial_r \hat B \right) \partial_r V_{rm} \nonumber\\
&&+\left[ e^{-\hat B} \left(m- e\overline{g}\,{\cal A}_{\varphi} \right)^2 - (\partial_r \hat A)^2 - \frac12 \partial_r \hat A \partial_r \hat B - \partial_r^2 \hat A -\frac12 \partial_r^2 \hat B \right] V_{rm} \nonumber\\
&&+ i e^{-\hat B} \left[ \left( \partial_r \hat B - \partial_r
\hat A \right) \left( m - e\overline{g}\, {\cal A}_{\varphi}
\right) +2 e\overline{g}\,
\partial_r {\cal A}_{\varphi} \right] V_{\varphi m} \label{EOMVr}
\eea and \bea e^{-\hat A} M_{m}^2 V_{\varphi m} &=& -\partial_r^2
V_{\varphi m} -  \left( 2 \partial_r\hat A - \frac{\partial_r \hat
B}{2} \right) \partial_r V_{\varphi m}
+ e^{-\hat B} \left(m- e\overline{g}\,{\cal A}_{\varphi} \right)^2 V_{\varphi m} \nonumber\\
&&- i \left[ \left( \partial_r \hat B - \partial_r \hat A \right)
\left( m - e\overline{g}\, {\cal A}_{\varphi} \right) +2
e\overline{g}\,
\partial_r {\cal A}_{\varphi} \right] V_{r m}, \label{EOMVvarphi}
\eea
where $M_{m}^2$ are the eigenvalues of $\eta^{\mu\nu} \partial_{\mu}\partial_{\nu}$.

At the same time, the variation leads to the following boundary conditions \cite{Nicolai:1984jg,massgaps}:
\be
\int d^4x dr \partial_r \left[e^{2\hat A + \hat B/2} \delta V^{\dagger}_{rm} \left(\partial_r - (\partial_r \hat A) \right) V_{rm} \right] = 0 \label{HCr}
\ee
and
\be
\int d^4x dr \partial_r \left[e^{2\hat A - \hat B/2} \delta V^{\dagger}_{\varphi m} \left(\partial_r - \frac12 (\partial_r \hat B) \right) V_{\varphi m} \right] = 0 \label{HCvarphi}
\ee
where we used $Tr(T^{I\dagger} T^J) = \delta^{IJ}$.  Notice that these are the weakest boundary conditions possible, which we must apply in order for the field equations to make sense mathematically.  In principle, with some physical motivation, we could apply stronger boundary conditions.  However, for the purposes of the stability analysis, since stronger boundary conditions would only eliminate modes from the physical spectrum, we prefer to remain as general as possible.

Our objective is then to solve these coupled linearized equations, together with their boundary conditions, in order to deduce the behaviour of the perturbations.

\subsection{The Schroedinger problem}
We proceed by transforming the system into a pair of coupled Schroedinger equations plus boundary conditions.
This is achieved by introducing the coordinate $\xi$, defined in ({\ref{theta}}), and the new variables:
\bea
V_{1m}(x,\xi) &\equiv& e^{\hat A/4 + \hat B/4} V_{\xi m}(x,\xi), \\
V_{2m}(x,\xi) &\equiv& e^{3\hat A/4 - \hat B/4} V_{\varphi m}(x,\xi)
\, . \eea The equations of motion (\ref{EOMVr}) and
(\ref{EOMVvarphi}) then become: \bea
M^2 V_1 &=& -V_1'' + \frac{1}{16} \left[ -4 \hat A'' - 4 \hat B'' + \left(\hat A' + \hat B'\right)^2 + 16 e^{\hat A - \hat B} \left(m - e\overline{g}\, {\cal A}_{\varphi} \right)^2 \right] V_1 \nonumber\\
&& + i e^{\hat A/2 - \hat B/2} \left[ \left(\hat B' - \hat A'
\right) \left( m - e\overline{g}\, {\cal A}_{\varphi} \right) + 2
e \overline{g}\,{\cal A}_{\varphi}' \right] V_2 \eea and \bea
M^2 V_2 &=&  -V_2'' + \frac{1}{16} \left[ 12 \hat A'' - 4 \hat B'' + \left(3 \hat A' - \hat B'\right)^2 + 16 e^{\hat A - \hat B} \left(m - e\overline{g}\, {\cal A}_{\varphi} \right)^2 \right] V_2 \nonumber\\
&&- i e^{\hat A/2 - \hat B/2} \left[ \left(\hat B' - \hat A'
\right) \left( m - e\overline{g}\, {\cal A}_{\varphi} \right) + 2
e\overline{g}\, {\cal A}_{\varphi}' \right] V_1\, , \eea where $'
\equiv
\partial_{\xi}$ and we have suppressed the index $m$.  In other
words, the system can then be described in the following way:
\be \left( \bacc -\partial_{\xi}^2 + U_1(\xi) & i C(\xi) \\
-i C(\xi) & -\partial_{\xi}^2 + U_2(\xi) \ea \right) \left( \bacc V_1(x,\xi) \\ V_2(x,\xi) \ea \right) =  M^2 \left( \bacc V_1(x,\xi) \\ V_2(x,\xi) \ea \right) \, , \label{coupledschro}
\ee
where the Schroedinger potentials are given by
\bea
U_1 &\equiv& \frac{1}{16} \left[ -4 \hat A'' - 4 \hat B'' + \left(\hat A' + \hat B'\right)^2 + 16 e^{\hat A - \hat B} \left(m - e\overline{g}\, {\cal A}_{\varphi} \right)^2 \right] \,, \\
U_2 &\equiv& \frac{1}{16} \left[ 12 \hat A'' - 4 \hat B'' +
\left(3 \hat A' - \hat B'\right)^2 + 16 e^{\hat A - \hat B}
\left(m - e\overline{g}\, {\cal A}_{\varphi} \right)^2 \right] \,
, \eea and the coupling function is \be C \equiv e^{\hat A/2 -
\hat B/2} \left[ \left(\hat B' - \hat A' \right) \left( m -
e\overline{g}\, {\cal A}_{\varphi} \right) +2 e\overline{g}\,
{\cal A}_{\varphi}' \right] \,. \ee

At this stage we can observe that for the conical solution
(\ref{GGPsolution}) the two Schroedinger potentials are degenerate:
\be U_1 = U_2 \equiv U \, . \ee It is therefore straightforward to
diagonalize the system.  Indeed, transforming into the basis:
\be
V_{\pm}(x,\xi) \equiv \frac{1}{\sqrt{2}} \left( V_1(x,\xi) \pm i V_2(x,\xi) \right)
\label{+-basis}
\ee
the matrix equation (\ref{coupledschro}) becomes:
\be \left( \bacc -\partial_{\xi}^2 + U(\xi) + C(\xi) & 0 \\
0 & -\partial_{\xi}^2 + U(\xi)- C(\xi)\ea \right) \left( \bacc V_+(x,\xi) \\ V_-(x,\xi) \ea \right) =  M^2 \left( \bacc V_+(x,\xi) \\ V_-(x,\xi) \ea \right) \, . \label{decoupledschro}
\ee

Having decoupled the equations, we should now consider the boundary
conditions in terms of the new basis (\ref{+-basis}). The sum and
difference of (\ref{HCr},\ref{HCvarphi}) lead to the following
constraints: \be \int d^4x d\xi \partial_{\xi} \left[ \delta
V_{+}^{\dagger} \left( \partial_{\xi} - \frac14 \hat B' \right) V_+
+  \delta V_{-}^{\dagger} \left( \partial_{\xi} - \frac14 \hat B'
\right) V_- \right] = 0 \label{HC+} \ee and \be \int d^4x d\xi \partial_{\xi}
\left[ \delta V_{+}^{\dagger} \left( \partial_{\xi} - \frac14 \hat
B' \right) V_- +  \delta V_{-}^{\dagger} \left( \partial_{\xi} -
\frac14 \hat B' \right) V_+ \right] = 0 \, . \label{HC-} \ee Here we have used
that $\hat A' \rightarrow 0$ at the boundaries, as can be seen from
(\ref{GGPsolution}) and (\ref{GGPsolution2}).  Notice that since the
dynamics of $V_+$ and $V_-$ are decoupled, we can consider both them and their
variations to be independent.  Therefore, choosing first $V_-(x,\xi) = 0$ and $\delta V_-(x,\xi)=0$, and then $V_+(x,\xi) = 0$ and $\delta V_+(x,\xi)=0$, we see that the boundary conditions
can be equivalently expressed as\footnote{Indeed, $(V_+(x,\xi),0)$ and $(0,V_-(x,\xi))$ are both well-defined solutions to the two-by-two Schroedinger system (\ref{decoupledschro}).  Then (\ref{HC1}) and (\ref{HC2}) are both necessary and sufficient boundary conditions.  We also checked explicitly that (\ref{HC+},\ref{HC-}) and (\ref{HC1},\ref{HC2}) are equivalent for our final solutions.} \be \int d^4x d\xi
\partial_{\xi} \left[ \delta V_{+}^{\dagger} \left( \partial_{\xi} -
\frac14 \hat B' \right) V_+ \right] = 0 \label{HC1}\ee and \be \int d^4x d\xi
\partial_{\xi} \left[ \delta V_{-}^{\dagger} \left( \partial_{\xi} -
\frac14 \hat B' \right) V_- \right] = 0 \, . \label{HC2}\ee In fact, these
conditions ensure that the Hamiltonians in the Schroedinger
equations (\ref{decoupledschro}) are Hermitian, and thus have real
eigenvalues and orthonormal sets of eigenfunctions.  We shall
therefore refer to them as Hermiticity Conditions (HCs).

Finally, we recall that in order to derive the physical spectrum of
the perturbations we must also impose the additional constraint of a
finite kinetic term in (\ref{b2action}). Expressed in term of the
decoupled fluctuations (\ref{+-basis}), this implies that: \be
-\frac12\int d^4x d\xi \left[ \partial_{\mu} V^{\dagger}_+
\partial^{\mu} V_+ \right] < \infty \label{NCV+} \ee and \be
-\frac12\int d^4x d\xi \left[ \partial_{\mu} V^{\dagger}_-
\partial^{\mu} V_- \right] < \infty \, . \ee

\subsection{The spectrum}
Having decoupled the equations in terms of the dynamical fields
$V_{+m}(x,\xi)$ and $V_{-m}(x,\xi)$ (restoring momentarily the index
$m$), let us decompose these fields in the standard Kaluza-Klein
way: \bea
V_{+m}(x,\xi) &=& {\cal V}_{+m}(x) \psi_{+m}(\xi) \, ,\nonumber\\
V_{-m}(x,\xi) &=& {\cal V}_{-m}(x) \psi_{-m}(\xi) \, .
\eea
Now, for $\psi_+$ the Schroedinger potential can be written explicitly as\footnote{Note that here we have discarded the delta-function contributions to the potential, since they are dominated by stronger singularities \cite{massgaps}.}:
\be U(\xi)+C(\xi)=U_0+u\cot^2\left(\frac{\xi}{r_0}\right)
+\overline{u}\tan^2\left(\frac{\xi}{r_0}\right),\label{SchroedingerVGGP}\ee
where
\bea
&& r_0^2U_0 \equiv \frac12 + 2m\omega - 2(m-N)\bar\omega + 2m(m-N)\omega\bar\omega,\nonumber \\
&& r_0^2u\equiv \frac34 + m^2 \omega^2 + 2m\omega,\quad r_0^2 \overline{u}\equiv
\frac34 + (m-N)^2 \bar\omega^2 - 2(m-N)\bar\omega\,,
\eea
and
\be
\omega\equiv(1-\delta/2\pi)^{-1}, \qquad
\bar\omega\equiv(1-\overline{\delta}/2\pi)^{-1}. \label{omegas}
\ee
Moreover, the HC reduces to:
\be \lim_{\xi\rightarrow
\overline{\xi}}\psi_+^*\left(-\partial_{\xi}+\frac{1}{2}\frac{1}{\xi-\overline{\xi}}
\right)\psi_+
-\lim_{\xi\rightarrow
0}\psi_+^*\left(-\partial_{\xi}+\frac{1}{2\xi}\right)\psi_+=0
\label{HC}\ee
and the finiteness of the kinetic energy (\ref{NCV+}) becomes simply the normalizability condition (NC) on the wavefunction:
\be
\int d\xi |\psi_+|^2 < \infty \, . \label{NC}
\ee

The potential, HC and NC for $\psi_-$ are identical to those for $\psi_+$, but replacing $m\rightarrow -m$ and $N \rightarrow -N$.

The problem is now of exactly the same form as that treated in
\cite{massgaps}, where the spectrum for gauge field and fermion
fluctuations was derived.  We can therefore follow the same steps
made there.  The Schroedinger equation can be transformed into the
hypergeometric equation:
\be z(1-z)\partial_z^2y+\left[c-(a+b+1)z\right]\partial_zy-ab
y=0\label{hyper},\ee
by defining
\be z=\cos^2\left(\frac{\xi}{r_0}\right),\qquad
\psi=z^{\gamma}\left(1-z\right)^{\beta}y(z),\ee
and,\phantom{b}for $\psi=\psi_+$\phantom{e},
\bea
\beta &\equiv& \frac{1}{4}\left(3+2m\omega\right),\,\,\gamma \equiv \frac{1}{4}\left[3-2(m-N)\bar\omega\right],
\,\, c\equiv 2-(m-N)\bar\omega, \nonumber \\
a&\equiv&\frac{1}{2}\left\{3+m\omega - (m-N)\bar\omega
+ \sqrt{r_0^2M^2+1+\left[m\omega-(m-N)\bar\omega\right]^2}\right\},\nonumber\\
b &\equiv&\frac{1}{2}\left\{3+m\omega - (m-N)\bar\omega
- \sqrt{r_0^2M^2+1+\left[m\omega-(m-N)\bar\omega\right]^2}\right\}.
\label{gbetaabcV}\eea
The solution can then be expressed in terms of Gauss's hypergeometric function, $F$.  For $c \neq 1$ the two linearly independent solutions are:
\be y_1(z)\equiv F(a,b,c,z), \quad y_2(z)\equiv z^{1-c}
F(a+1-c,b+1-c, 2-c, z), \label{y12}\ee
and so the general solution to the Schroedinger equation is
\be\psi=K_1\psi_1 + K_2 \psi_2,\label{cnot1}\ee
with
\be \psi_i\equiv z^{\gamma}(1-z)^{\beta}y_i \label{psii} \ee
and $K_{1,2}$ the integration constants. For $c=1$ we have
$\psi_1=\psi_2$ but we can construct a linearly independent solution
using the Wronskian method and the general solution reads
\be\psi=K_1\psi_1 + K_2
\psi_1\int^{\xi}\frac{d\xi'}{\psi_1^2(\xi')}.\label{wronskian}\ee

We must now impose the NC (\ref{NC}) and HC (\ref{HC}) to select the
physical modes. The explicit calculations are given in Appendix C of
\cite{massgaps}, and so we do not repeat them here.  The final
result is the following.

The wavefunctions for $\psi_+$ are:
\bea \psi_+&\propto&  z^{\gamma}(1-z)^{\beta}F(a,b,c,z), \quad for\quad m\leq
N + 1/\bar\omega,\label{psi1}\\
\psi_+&\propto&  z^{\gamma +1-c}(1-z)^{\beta} F(a+1-c,b+1-c, 2-c,
z), \quad for\quad m>N + 1/\bar\omega.\label{psi2} \eea
The corresponding squared masses are:
\begin{itemize}
\item For  $ m \leq -1/\omega$ and $m \leq N + 1/\bar\omega$
\be
M^2=\frac{4}{r_0^2}\left\{n(n+1)-\left(n+\frac{1}{2}\right)\left[
m\omega+(m-N)\bar\omega\right] + m(m-N)\omega\bar\omega \right\}.\label{M1}\ee
\item For  $-1/\omega < m \leq N + 1/\bar\omega$
\be
M^2=\frac{4}{r_0^2}\left\{\left(n+\frac32\right)^2 - \frac14 +\left(n+\frac{3}{2}\right)\left[
m\omega-(m-N)\bar\omega \right]\right\}. \phantom{0000000000} \label{M2}\ee
\item For $N + 1/\bar\omega < m \leq -1/\omega$
\be
M^2=\frac{4}{r_0^2}\left\{n(n-1)-\left(n-\frac{1}{2}\right)\left[
m\omega - (m-N)\bar\omega\right]\right\}. \phantom{000000000000000} \label{M3}\ee
\item For  $m>-1/\omega$ and $m > N + 1/\bar\omega$
\be
M^2=\frac{4}{r_0^2}\left\{n(n+1)+\left(n+\frac{1}{2}\right)\left[
m\omega+(m-N)\bar\omega\right] + m(m-N)\omega\bar\omega \right\}.\label{M4}
\ee
\end{itemize}
Here $n=0,1,2,\dots$.  The masses given in (\ref{M1}) and (\ref{M2})
correspond to the wave function (\ref{psi1}) whereas the masses
given in (\ref{M3}) and (\ref{M4}) correspond to the wave function
(\ref{psi2}). The spectrum for $\psi_-$ can be obtained from that
above simply by transforming $m \rightarrow -m$ and $N \rightarrow
-N$.

\section{Stability Analysis} \label{S:stability}
In the preceding section we have analytically derived the complete
Kaluza-Klein spectrum for the scalar fluctuations of the 6D gauge
field, for directions in the Lie Algebra of ${\cal G}$ orthogonal to
the background monopole.  We are now ready to analyze the stability
of these fluctuations.  Thanks to the canonical kinetic terms, the
question amounts to whether or not there are any tachyonic modes.

\subsection{The sphere case} \label{spherestability}
Let us first describe what happens in the well-known sphere
case. To this end, it is illuminating to write the
spectrum (\ref{M1})-(\ref{M4}) in the following way: \be M^2 =
\frac{4}{r_0^2} \left[l(l+1) - \left(\frac{P}{2}\right)^2 \right]
\label{Msq} \ee where, for $\psi=\psi_+$, $P =
m\omega - (m-N)\bar\omega$, $l = k + |1 + P/2|$ and we
have the following definition of $k$ in the two cases $P>-2$ and
$P\leq-2$:

\begin{description}

\item[{\Large $P > -2$}]:
\begin{itemize}
\item For  $ m \leq -1/\omega$
\be
k=n - m\omega -1 \geq 0 \label{k1}
\ee
\item For  $-1/\omega < m \leq N + 1/\bar\omega$
\be
k=n \geq 0
\ee
\item For $m > N + 1/\bar\omega$
\be
k=n+(m-N)\bar\omega - 1 > 0
\ee
\end{itemize}

\item[{\Large $P \leq -2$}]:

\begin{itemize}
\item For $m \leq N + 1/\bar\omega$
\be
k=n + 1 - (m-N)\bar\omega \geq 0
\ee
\item For $N + 1/\bar\omega < m \leq -1/\omega$
\be
k=n \geq 0
\ee
\item For  $m>-1/\omega$
\be
k=n+m\omega + 1 > 0\label{kf}
\ee
\end{itemize}
\end{description}
From here it is easy to confirm that in the sphere limit ($\omega,
\bar\omega \rightarrow 1$), the spectrum for $\psi_+$ reduces to the
expected form: \be {\mathcal R}^2 M^2 = l(l+1) - \left(\frac{N}{2}\right)^2 \,
\qquad \mbox{multiplicity}= 2l+1\,,\label{Nsphere} \ee where
${\mathcal R}=r_0/2$ represents the radius of the sphere, $l = k + |1 +
N/2|$ and $k$ is an integer which assumes all possible
non-negative values ($k=0,1,2,...$).  Recall that the
results for $\psi_-$ are obtained by taking $m\rightarrow -m$ and $N
\rightarrow -N$.  By
using this information it is easy to see that a necessary and
sufficient condition for the absence of tachyons in the sphere case
is simply \cite{Avramis:2005qt}
\be \left|N^I\right|\leq 1 \quad \mbox{for every} \,\,I,\label{spherecondition}\ee
where we have restored the Lie algebra index $I$. To derive Inequality (\ref{spherecondition}) one can use the fact that, in the
sphere case, $k=0$ is an allowed\footnote{This property does not
hold always for the conical-GGP solutions.} value of $k$ for every
$m$, as can be checked by means of the definitions
(\ref{k1})-(\ref{kf}).

In order to satisfy (\ref{spherecondition}) we must have that all
the charges $e^I$ corresponding to our sector (\ref{oursector})
assume just one value, up to their sign. Moreover, we must have that
the absolute value of all the charges in the hypermatter sector, $|e^i|$, is
not smaller than $|e^I|$. These are consequences of the Dirac
quantization condition (\ref{DiracQ}). If we embed the background
monopole in an Abelian factor of $\mathcal{G}$, these conditions are
obviously satisfied as $e^I=0$, for all $I$, in this case. However,
for an embedding in a non-Abelian factor of $\mathcal{G}$, these
conditions select only one possibility amongst the
anomaly free models presented in Table \ref{anomalyfree}: the $E_7 \times E_6 \times U(1)_R$ model, with the
monopole embedded in $E_6$ under which all hypermultiplets are singlets
\cite{seifanomaly,Avramis:2005qt,Avramis:2005hc}\footnote{There are other non-trivial examples.  For instance, amongst the numerous models given in \cite{Suzuki:2005vu}, if we again take the monopole to lie in a non-Abelian factor, stability selects the $SU(2) \times U(1)_R$ model with hyperino representation as follows: seven hyperinos transforming
as a ${\bf 3}$ of $SU(2)$, two as a ${\bf 5}$ and thirty-one as a
${\bf 7}$ \cite{alberto}.}.

\subsection{The conical-GGP case} \label{stGGPcase}
Our purpose is now to see whether or not Condition
(\ref{spherecondition}) is valid also for the conical-GGP solutions.
To this end one can analyze directly the explicit expressions for the
spectrum given in (\ref{M1})-(\ref{M4}) and simply study the
inequality $M^2\geq 0$ for those four expressions, which are valid
in four different ranges of $m$. After a long but straightforward
computation, the following results emerge\footnote{Recall that we have not imposed the orbifold boundary conditions on the present spectrum.  By eliminating some of the modes the orbifolding could relax some of the stability conditions.  The conditions we state are, however, certainly sufficient also in this case.}. A sufficient condition
for the absence of tachyons in the conical-GGP solution is
Constraint (\ref{spherecondition}). Thus the warping and brane defects do not introduce new instabilities.  Meanwhile, when both the tensions are non-negative ($T\geq 0$ and $\overline{T}\geq 0$), Constraint
(\ref{spherecondition}) is also a necessary condition for the
absence of tachyons. In this case we have exactly the same situation
as in the sphere case, so positive tension branes have no effect on stability at all. However, when at least one tension is
negative ($T<0$ and/or $\overline{T}<0$), we can relax that
constraint if the absolute value of a negative tension is large
enough. The latter statement can be proved by analyzing the
following special cases.

\begin{description}

\item[(i)] \hspace{0.7cm}$T<0$ and $\overline{T}=0$, that is $\omega<1$ and $\overline{\omega}=1$.
This case corresponds to a solution with just one conical defect and
a non-trivial warping. Here a necessary and sufficient condition for
the absence of tachyons is
\be \left|N^I\right|\leq 1 +\frac{1}{3\omega} \quad \mbox{for every} \,\,I.\label{i}\ee
\item[(ii)] \hspace{0.7cm}$T=\overline{T}<0$, that is $\omega=\overline{\omega}<1$.
This set up corresponds to the unwarped rugby ball compactification with negative deficit angles.
In this case a sufficient condition for the absence of tachyons is
\be \left|N^I\right|\leq \frac{4}{3\omega}\quad \mbox{for every} \,\,I.\label{ii}\ee

\end{description}
Some details of the derivation of (\ref{i}) and (\ref{ii}) are provided in Appendix \ref{detailsstab}.  Since $\omega$ appears in the denominator of (\ref{i}) and
(\ref{ii}), it is clear that we can render stable an arbitrary large
value of $|N^I|$ by choosing a small enough value of
$\omega$, that is by introducing a large negative tension brane. For example,
if we want to stabilize the value $|N^I|=2$, which is
unstable in the sphere case, we have to choose a deficit angle
$\delta\leq -4\pi$ in case (\ref{i}), and $\delta \leq -\pi$ in case (\ref{ii}).

The main conclusion is that the conical-GGP solution is a stable
solution for all 6D gauged supergravities, if we allow arbitrarily
negative brane tensions. However, if we require that there are only
non-negative tensions, the stability of the system exactly selects
the same models as in the sphere case (with the additional topological constraint that the monopole cannot be embedded in $U(1)_R$ -- see the end of subsection \ref{S:geometry}).

\section{Conclusions}
We have studied the stability of axi-symmetric brane world compactifications (conical-GGP solutions) in anomaly free, chiral, gauged 6D supergravity.  Anomaly freedom is a central principle of quantum physics, and in six dimensional supergravity it places restrictions on the possible matter supermultiplets that can be present.  Indeed, this lack of arbitrariness can be considered as one of the theory's most attractive features.  Meanwhile, we remark that the central results of our work also apply to more general situations.  For example, they are relevant for any 6D gauged supergravity (anomalous or otherwise) which has non-Abelian gauge groups, and also for the Yang-Mills extensions to the non-supersymmetric models of \cite{Carroll:2003db}.

We began by considering the various types of bulk and brane scalar fluctuations that are present in the model, and in particular how they decouple at the bilinear level.  We chose to project out the brane bending modes by placing the branes at orbifold fixed points.  Then, the Salam-Sezgin sector ({\it i.e.} those scalars fluctuations descending from the supergravity-tensor multiplet and the $U(1)$ gauge multiplet in the direction of the background monopole) was considered for the axially-symmetric perturbations in \cite{kicking}, and no instabilities were found.

The remaining sectors are then the hyperscalar fluctuations, and those scalar fluctuations of the gauge fields which are orthogonal to the monopole background in the Lie algebra of ${\cal G}$.  The hyperscalars fluctuations can immediately be seen to have positive squared-masses in the 4D effective theory.  Therefore, the main efforts in the present article were directed towards the scalar fluctuations of the said gauge fields, the sector which normally harbours instabilities.

We used the light-cone gauge to derive the bilinear action for these fluctuations \cite{lightcone}. In this way we obtained the linearized equations of motion, and found them to be a pair of coupled, second-order ODEs.  We transformed them into a pair of Schroedinger equations, and found that it is possible to decouple them.  The problem then reduces to the same form as that treated in \cite{massgaps}, and we analytically derived the full spectra.  The exact results that we obtained enabled us to draw both expected and surprising results, with regards to the stability of the compactifications.

As was observed long ago for the sphere-monopole compactification of these theories \cite{seifinstabilities}, in general we find a tachyonic instability in the scalar fluctuations of the gauge fields that are charged under the monopole background.  In the case of the sphere, the necessary and sufficient condition for the absence of tachyons can be written as $|N^I| \leq 1$, where $N^I$ are the integer monopole numbers carried by each gauge field \cite{Avramis:2005qt}.  If we embed the monopole background in an Abelian factor of the gauge group, then the compactification is stable.  However, if the monopole happens to lie in a non-Abelian factor of the gauge group, then generically the compactification is unstable.  For example, amongst the anomaly free models described in Table \ref{anomalyfree}, only one fulfills the necessary condition: the $E_7 \times E_6 \times U(1)_R$ model with the monopole embedded in the hidden $E_6$ \cite{seifanomaly,Avramis:2005qt,Avramis:2005hc}.

We find that the same condition holds for the conical-GGP solutions that contain positive tension branes only.  It is also a sufficient condition for models which incorporate negative tension branes.  In other words, a GGP compactification is stable if its sphere limit is stable.  Furthermore, with positive tension branes only, a GGP compactification is unstable if its sphere limit is unstable.  However, it becomes possible to relax the constraint by incorporating large, negative tensions.  This seems remarkable, given that negative tension branes are usually associated with instabilities rather than stability.

Meanwhile, the simplest way to obtain a stable conical-GGP solution appears to be to embed the monopole in an Abelian factor of the gauge group.  Notice that, at least for the most elegant anomaly free models such as those in Table \ref{anomalyfree}, there is typically only one Abelian gauge factor available, which corresponds to the $U(1)_R$ gauged R-symmetry.  It is known, however, from the Dirac quantization condition, that embedding the monopole in the $U(1)_R$ again leads to the requirement of at least one negative tension brane \cite{GGP, generalsoln}!

We should note, though, that there is actually a host of anomaly free models with extra drone $U(1)$'s \cite{Avramis:2005hc,Suzuki:2005vu}.  For a stable model with only positive tension branes, therefore, we must turn to one of these or to one of the more ``miracolous'' models discussed at the end of Subsection \ref{spherestability}.

For these reasons, we have also discussed in detail the geometry induced by negative tension branes.  The associated negative deficit angles give rise to so-called {\it saddle-cones}.  As noted, the orbifolding projects out the brane-embedding fields, which are usually a source of instability for negative tension branes.  We mention here that, furthermore, the orbifolding gives rise to a chiral spectrum for the bulk fermion zero modes.  Whilst models with positive tension branes give rise to a chiral spectrum even without the orbifold projection, those which include negative tension branes in general include zero modes of both chirality.  This can be observed by considering the spectrum found in \cite{massgaps}. In addition we note that the main conclusion of Ref. \cite{massgaps} -- that the Kaluza-Klein mass scale
and that of the internal
volume  can decouple in the presence of conical defects -- also holds for the
present mass spectrum in the case of negative tension branes.

In summary, we find that the conical-GGP solutions with positive tension branes are stable only for very limited classes of anomaly free theories and monopole embeddings.  Such models, therefore, as used for example in the Supersymmetric Large Extra Dimension Scenario, are almost unique in character.  Meanwhile, somewhat surprisingly, negative tension branes seem to allow for stability in a much wider class of models.  It would certainly be interesting to obtain some physical intuition as to how the negative tension branes render unstable sphere compactifications stable.

Finally, given that the generic model is unstable, the big question is: where is the instability taking it to?  Here we comment that since the tachyonic mass has its origin in the internal part of the 6D gauge kinetic term, which is semi-positive definite, we expect it to be stabilized at the quartic level.   Moreover, the tachyonic masses are found in non-axially symmetric modes, so we might ask: is there a stable non-axially symmetric brane world solution to 6D gauged supergravity?

\vspace{1cm}

{\bf Acknowledgments.}   We would especially like to thank Sergio Azevedo, Mario Mazzioni, H. Chacham and R.W Nunez for kindly providing us with Figure 2.  It is also our pleasure to thank Cliff Burgess, Andrew Frey, Sean Hartnoll, Kazuya Koyama, Antonios Papazoglou, Riccardo Rattazzi and Andrew Tolley for interesting and useful communications. The work of A.S is supported in part by the Swiss National Science Foundation.  S.L.P thanks the High Energy Theory Group at McGill University and the Perimeter Institute for their hospitality at the last stages of this work.

\newpage
\appendix

{\Large \bf Appendix}

\section{Bilinear Action in the Light-Cone Gauge} \label{A:gaugefixing}
In order to derive the bilinear action for the scalar fluctuations
of the gauge field orthogonal to the monopole background, we use the
results of \cite{lightcone}.  In that reference, a formalism was
developed to analyze the spectrum of small perturbations about
arbitrary solutions of Einstein, Yang-Mills and scalar systems,
using the light-cone gauge.\footnote{Some discussions of the light
cone gauge in field theory are given in
\cite{Randjbar-Daemi:1984ap,Randjbar-Daemi:1984fs,lightcone}.}  For
a warped background solution, with their scalars inactive, the model
turns out to be identical to ours, up to the latter's presence of
the dilaton.\footnote{Our model also contains the Kalb-Ramond field and hyperscalars,
but again, for the background and fluctuations of interest, these
sectors does not contribute.}  In this appendix, we show how our
model can in fact be transformed into exactly the system treated in
\cite{lightcone}.

The bilinear action for the sector of interest about the background
(\ref{axisymmetric}) will have contributions only from the gauge
kinetic term:
\be S = -\frac{1}{4}\int d^6X \sqrt{-G} \left[e^{\kappa\sigma/2}
G^{MN} G^{PQ} Tr\left( {F}_{MP} {F}_{NQ} \right) \right] \ee
 We can make a
conformal transformation to absorb the dilaton by defining $\hat
G_{MN} = e^{\phi} G_{MN}$, where we recall that $\phi \equiv
\kappa\sigma/2$.  In the new frame, the action is identical to
that considered in \cite{lightcone}: \be S = -\frac{1}{4}\int d^6X \sqrt{-\hat G}
\left[ \hat G^{MN} \hat G^{PQ} Tr\left( {F}_{MP} {F}_{NQ} \right)
\right] \label{confaction} \ee

The conformal transformation, however, implies that our background metric becomes:
\be
ds^2 = e^{A+\phi} \eta_{\mu\nu} dx^{\mu} dx^{\nu} + e^{\phi} d\rho^2 + e^{B+\phi} d\varphi^2 \, ,
\ee
which differs from that considered in \cite{lightcone}, where $ds^2 = e^{A} \eta_{\mu\nu} dx^{\mu} dx^{\nu} + d\rho^2 + e^{B} d\varphi^2$.  We therefore make a coordinate tranformation $e^{\phi/2} d\rho \equiv dr$, so that our metric can be written as:
\be
ds^2 = e^{\hat A} \eta_{\mu\nu} dx^{\mu} dx^{\nu} + dr^2 + e^{\hat B} d\varphi^2 \, ,
\ee
with $\hat A \equiv A + \phi$ and $\hat B \equiv B + \phi$.  The action ({\ref{confaction}}) of course remains invariant under the change of coordinates.

After these tricks, we can follow exactly the same steps performed
in \cite{lightcone} to remove the gauge degrees of freedom and
obtain the dynamics of the physical fields.  We expand to bilinear
order, transform to light-cone coordinates, fix to the light-cone
gauge and eliminate the redundant degrees of freedom using their
equations of motion.  The final result, for the spin-0 fields $V_r$ and $V_{\varphi}$ orthogonal to the background monopole, is then the following
bilinear action: \bea S_{2}(V,V) &=& -\frac12 \int d^6X \,
\sqrt{-\hat G}\,\,Tr \left[\partial_{\mu} V_i \partial^{\mu} V^i +
D_i V_j D^i V^j - 2 (\partial_r \hat A)^2 V_r^2  \right. \nonumber
\\ && \phantom{000000000}  \left. - 2 (\partial_r \hat A) V_r D_iV^i
 + \hat R_{ij}V^iV^j + 2 \overline{g}\,F_{ij}V^i \times V^j \right] \, ,
\eea where now $\hat G_{MN}$, $\sigma$ and $F_{ij}$ refer to the
background fields and $V_i$ to the fluctuations. The index $i$ runs
over $r, \varphi$ and all indices are raised and lowered with $\hat
G_{MN}$.

\section{Details of Stability Analysis} \label{detailsstab}
In this appendix we give some intermediate steps to obtain the results of Subsection \ref{stGGPcase}, concerning
the stability analysis for the conical-GGP solutions in the
presence of negative tension branes. Indeed, only in the presence of at least one negative tension brane our results differ from the stability constraint
given in (\ref{spherecondition}) that is valid in the sphere case.  Therefore, as in Subsection \ref{stGGPcase}, here we focus on the following special cases

\begin{description}

\item[(i)] \hspace{0.7cm}$T<0$ and $\overline{T}=0$, that is $\omega<1$ and $\overline{\omega}=1$,

\item[(ii)] \hspace{0.7cm}$T=\overline{T}<0$, that is $\omega=\overline{\omega}<1$,

\end{description}
whose analysis is enough to obtain the main results of the present paper. More precisely, in the following  we show how to obtain Constraints (\ref{i}) and (\ref{ii}).

Let us first consider Case (i). The mass squared spectrum for the scalar sector (\ref{oursector}), which we are interested in, is given in Equations
(\ref{M1})-(\ref{M4}), for the $\psi_+$ wave function\footnote{We recall that the spectrum for $\psi_-$ can be obtained by transforming $m\rightarrow -m$ and
$N\rightarrow -N$.}, and one has to take into account all of them to perform a complete analysis. However, here we only consider Equation (\ref{M2})
because the analysis of (\ref{M1}), (\ref{M3}) and (\ref{M4}) is analogous. By using Ansatz (i), Equation (\ref{M2}) becomes
\be \lambda = \left(n+\frac32\right)^2 - \frac14 +\left(n+\frac{3}{2}\right)\left[
m\omega-(m-N) \right],
\ee
where $\lambda\equiv r_0^2M^2/4$. We recall that (\ref{M2}) is valid for $-1/\omega < m \leq N + 1/\bar\omega$, which, for $\overline{\omega}=1$, becomes
\be m\omega>-1 \quad \mbox{and} \quad m-N\leq 1. \label{constraintM2} \ee
Now we observe that
\bea \lambda &=& n^2+ \left[3+m\omega -(m-N)\right]n +2 +\frac32 \left[m\omega - (m-N)\right]\nonumber \\
&&\geq 2 +\frac32 \left[m\omega - (m-N)\right]=\lambda_0, \eea
where $\lambda_0$ is $\lambda$ evaluated at $n=0$ and we used $3+m\omega-(m-N)>1$, which is a consequence of (\ref{constraintM2}). Therefore, we have $\lambda\geq 0$
if $\lambda_0\geq 0$, namely if
\be m\omega \geq m-N-\frac43. \label{effconstraint} \ee
Now we observe that, for $m-N\leq 0$, Constraint (\ref{effconstraint}) is always satisfied because of (\ref{constraintM2}). Therefore, tachyons can only be present
in the case\footnote{We observe that Constraints (\ref{constraintM2}) forbid  $m> N+1$.} $m=N+1$ and by plugging this value of $m$ into (\ref{effconstraint}) we obtain
\be N\geq -1-\frac{1}{3\omega}. \label{N+} \ee
The corresponding constraint for $\psi_-$ can be obtained by transforming $N\rightarrow -N$ in (\ref{N+}); this leads to
$N\leq 1+1/(3\omega)$, which, together with (\ref{N+}),
gives exactly (\ref{i}). By using a similar method, we further checked that (\ref{i}) is a necessary and sufficient condition for $\lambda\geq 0$ for the full spectrum including also (\ref{M1}), (\ref{M3})
and (\ref{M4}).

Let us now consider Case (ii). Again we present only the analysis of Equation (\ref{M2}), which, for $\omega=\overline{\omega}$, is
\be
\lambda=\left(n+\frac32\right)^2 - \frac14 +\left(n+\frac{3}{2}\right)
N\omega.  \ee
The mass squared given in (\ref{M2}) is valid for $-1/\omega < m \leq N + 1/\bar\omega$, which reduces to
\be m\omega>-1 \quad \mbox{and} \quad (m-N)\omega\leq 1\label{newconstraintM2}\ee
in the case $\omega=\overline{\omega}$.
As we did in Case (i), we observe that
\be \lambda = n^2+ \left(3+N\omega\right)n +2 +\frac32 N\omega \,\geq\, 2 +\frac32 N\omega=\lambda_0, \ee
where we used $3+N\omega>1$, which is a consequence of (\ref{newconstraintM2}). So a sufficient condition for $\lambda\geq 0 $ is
\be N\geq -\frac{4}{3\omega} \label{N+2} \ee
and, by also taking into account the corresponding constraint for $\psi_-$ ($N\leq 4/(3\omega)$), we obtain (\ref{ii}). Analogously we checked that (\ref{ii})
is a sufficient condition for $\lambda\geq0$ for (\ref{M1}), (\ref{M3})
and (\ref{M4}) as well.

\end{document}